\documentclass[aps,preprint,showpacs]{revtex4-1}
\usepackage{graphicx}
\usepackage{bm}
\usepackage{amsmath, upgreek}
\begin{document}
\title{Drift of suspended ferromagnetic particles due to the Magnus effect}
\author{S.~I.~Denisov}
\email{denisov@sumdu.edu.ua}
\author{B.~O.~Pedchenko}
\affiliation{Sumy State University, 2 Rimsky-Korsakov Street, UA-40007  Sumy, Ukraine}


\begin{abstract}
 A minimal system of equations is introduced and applied to study the drift motion of ferromagnetic particles suspended in a viscous fluid and subjected to a time-periodic driving force and a nonuniformly rotating magnetic field. It is demonstrated that the synchronized translational and rotational oscillations of these particles are accompanied by their drift in a preferred direction, which occurs under the action of the Magnus force. We calculate both analytically and numerically the drift velocity of particles characterized by single-domain cores and nonmagnetic shells and show that there are two types of drift, unidirectional and bidirectional, which can be realized in suspensions composed of particles with different core-shell ratios. The possibility of using the effect of bidirectional drift for the separation of core-shell particles in suspensions is also discussed.
\end{abstract}
\pacs{47.65.Cb, 75.75.Jn, 82.70.-y}

\maketitle

\textit{Introduction.}---The Magnus effect, i.e., difference between the trajectories of rotating and non-rotating bodies moving through a medium, has a long and rich history. Newton was probably the first who described this phenomenon for a tennis ball \cite{Newt}. Then Robins \cite{Rob} and much later Magnus \cite{Magn} experimentally investigated the influence of rotation of musket and cannon balls on their trajectories, and the first theoretical explanation was given by Rayleigh \cite{Rayl}. Nowadays the Magnus effect is used to describe the motion of rotating bodies in sport \cite{Meht}, aeronautics \cite{Seif} and planet formation \cite{YaKi, Forb}, to name a few. This effect is usually associated with the Magnus force, which acts on the body, accounts for its rotation, and is responsible for the difference between the trajectories. Note also that a similar force (but of different origin) plays an important role in dynamics of vortices in superconductors \cite{NoVi, AoTh}, superfluids \cite{Son}, and magnets \cite{ThSt}.

Because the Magnus force depends on many factors, including the dynamical characteristics of the body, shape and roughness of the interface and properties of the surrounding medium, its theoretical determination is a complex problem. But if the solution of this problem and the viscous drag force and torque acting on the body are known, then studying the body motion, both translational and rotational, is greatly simplified since the dynamics of the host medium can be excluded from consideration. In particular, according to \cite{RuKe}, such a simplification is possible for spherical bodies characterized by smooth surfaces and small Reynolds numbers. Within this approximation, the rotational properties of ferromagnetic particles suspended in a viscous fluid and subjected to a uniform time-dependent magnetic field have already been thoroughly studied (see, e.g., Refs.~\cite{LDRB, UsUs} and references therein). At the same time, to the best of our knowledge, the role of the Magnus force in transport properties of these particles has never been discussed before. The aim of this paper is to show that the Magnus effect can be used to generate their directed transport.

\textit{Definitions and background.}---Our approach is based on the simplest system of equations that describes the translational and rotational motion of ferromagnetic particles in a viscous fluid and that accounts for the Magnus effect. We consider the spherical core-shell particles each of which is characterized by a single-domain core of radius $a$ and magnetization $\mathbf{M} = \mathbf{M}(t)$ and by a nonmagnetic shell of radius $b$ ($b\geq a$). We assume that the average distance between suspended particles is so large that their magnetic dipole-dipole and hydrodynamic interactions are negligibly small. It is also assumed that the particle dynamics occurs so slowly that the induced fluid flow is laminar and the inertial effects can be excluded from consideration. This means that the translational and rotational Reynolds numbers defined as $\mathrm{Re}_{t} = \rho b v_{m}/ \eta$ and $\mathrm{Re}_{r} = \rho b^{2} \omega_{m}/ \eta$, respectively, where $\rho$ and $\eta$ are the fluid density and dynamic viscosity, $v_{m} = \max{|\mathbf{v}|}$, $\omega_{m} = \max{|\boldsymbol{ \omega}|}$, and $\mathbf{v}$ and $\boldsymbol{ \omega}$ are the instantaneous translational and angular particle velocities, should be sufficiently small: $\mathrm{Re}_{t} \ll 1$ and $\mathrm{Re }_{r} \ll 1$. 

In our model, the translational motion of particles occurs under the action of three forces, namely, the frictional force $\mathbf{f}_{f}$, the Magnus lift force $\mathbf{f}_{l}$, and the driving force $\mathbf{f}_{d}$. With the above conditions, the friction force is determined by the Stokes law, $\mathbf{f}_{f} = -6\pi\eta b\mathbf{v}$, and the Magnus force is given by $\mathbf{f}_{l} = \pi\rho b^{3} \boldsymbol{ \omega} \times \mathbf{v}$ (the sign $\times$ denotes the vector product) \cite{RuKe}. Next we chose the driving force in the form $\mathbf{f}_{d} = f_{0} \sin{(\Omega t)} \mathbf{e}_{x}$, where $f_{0}$ and $\Omega$ are the amplitude and angular frequency of the force and $\mathbf{e}_{x}$ is the unit vector along the $x$ axis. Neglecting the inertial term, the equation for the translational motion of particles reduces to the force balance equation $\mathbf{f}_{f} + \mathbf{f}_{l} + \mathbf{f}_{d} = 0$, which can be solved with respect to the dimensionless particle velocity $\mathbf{u} = \mathbf{v}/v_{0}$ ($v_{0} = f_{0}/6\pi b\eta$) as
\begin{equation}
    \mathbf{u} = \frac{\mathbf{e}_{x} -
    \mathbf{e}_{x} \times \boldsymbol{\kappa} +
    \boldsymbol{\kappa} (\mathbf{e}_{x}\cdot \boldsymbol{\kappa})}
    {1 + \boldsymbol{\kappa}^{2}} \sin{(2\pi \tau)},
    \label{u}
\end{equation}
where $\boldsymbol{\kappa} = \mathrm{Re}_{r}\boldsymbol{\omega}/ 6\omega_{m}$ and $\tau = \Omega t/2\pi$ are the dimensionless angular velocity and time, respectively.

According to (\ref{u}), the rotational motion of particles characterized by the dimensionless angular velocity $\boldsymbol{\kappa}$ influences their translational motion. We next assume that the particle rotation is caused by a time-dependent magnetic field $\mathbf{H}$. The torque exerted on the particle magnetic moment $\boldsymbol{\mu} = 4\pi a^{3}\mathbf{M}/3$ by this field is given as $\mathbf{t}_{H} = \boldsymbol{\mu} \times \mathbf{H}$, and the frictional torque acting on the rotating particle is equal to $\mathbf{t}_{f} = -8\pi \eta b^{3} \boldsymbol{\omega}$. If the anisotropy magnetic field essentially exceeds the external field (this is a common situation), then the particle magnetic moment $\boldsymbol{\mu}$ can be considered as frozen into the material. In this case, the rotation of $\boldsymbol{\mu}$ is governed by the kinematic equation $d\boldsymbol{\mu}/dt = \boldsymbol{\omega} \times \boldsymbol{\mu}$, and the particle angular velocity at $\mathrm{Re}_{r} \ll 1$ satisfies the torque balance equation $\mathbf{t}_{H} + \mathbf{t}_{f} = 0$. Solving it for $\boldsymbol{ \omega}$ and substituting into the kinematic equation, we obtain the following equation for the unit magnetization vector $\mathbf{m} = \mathbf{M}/M$ ($M = |\mathbf{M}|$):
\begin{equation}
    \dot{\mathbf{m}} = -\alpha \mathbf{m}
    \times (\mathbf{m} \times \mathbf{h}).
    \label{eq_m}
\end{equation}
Here, the overdot denotes the derivative with respect to $\tau$, $\alpha = \alpha_{0} \nu^{3}$ is the dimensionless parameter associated with the inverse rotational relaxation time, $\alpha_{0} = \pi MH_{m}/3\eta \Omega$, $\nu = a/b \leq 1$ is the core-shell ratio, $H_{m} = \max{|\mathbf{H}|}$, and $\mathbf{h} = \mathbf{H}/H_{m}$.

Thus, the influence of rotation of particles on their translational motion, arising from the Magnus effect, can be investigated by analyzing the linear velocity (\ref{u}), in which the vector parameter $\boldsymbol{ \kappa}$  is determined by solving Eq.~(\ref{eq_m}) for $\mathbf{m}$ and calculating the angular velocity $\boldsymbol{ \omega} = (\alpha \Omega/2\pi) \mathbf{m} \times \mathbf{h}$. Here we consider the case when the magnetic field of a fixed magnitude ($|\mathbf{H}| = H_{m}$) rotates nonuniformly in the $xy$ plane and has the same periodic properties as the driving force. That is, we assume that $\mathbf{h} = \cos{\psi}\, \mathbf{e}_{x} + \sin{\psi}\, \mathbf{e}_{y}$, where the azimuthal angle of $\mathbf{h}$, $\psi = \psi(\tau + \phi/ 2\pi)$, is a given function of $\tau$ satisfying the condition $\psi|_{1/2 + \tau} = -\psi|_{\tau}$ (and so $\psi|_{1+ \tau} = \psi|_{\tau}$), and $\phi \in [0,2\pi]$ is the initial phase. Because in this field the steady-state solution of Eq.~(\ref{eq_m}), $\mathbf{m}_{ \mathrm{st}}$, lies in the $xy$ plane, it is reasonable to choose the magnetization vector $\mathbf{m}$ in the form $\mathbf{m} = \cos{ \varphi} \, \mathbf{e}_{x} + \sin{\varphi}\, \mathbf{e}_{y}$ [$\varphi = \varphi (\tau)$ is the azimuthal angle of $\mathbf{m}$]. In this case, using the relation $\mathbf{m} \times \mathbf{h} = \sin{\chi}\, \mathbf{e}_{z}$, where the lag angle $\chi = \chi(\tau)$ is defined as $\chi = \psi - \varphi$, Eq.~(\ref{eq_m}) reduces to the equation for the periodically driven overdamped pendulum
\begin{equation}
    \dot{\chi} + \alpha\sin{\chi} =\dot{\psi}.
    \label{eq_chi}
\end{equation}

Taking into account that the dimensionless angular velocity $\boldsymbol{\kappa}$ is expressed through the lag angle $\chi$ as $\boldsymbol{\kappa} = \gamma \sin{\chi} \, \mathbf{e}_{z}$, where $\gamma = \gamma_{0}\nu^{3}$ and $\gamma_{0} = \rho MH_{m} b^{2}/ 36\eta^{2}$, and neglecting the terms of the second order in $\mathrm{Re}_{r}$, from (\ref{u}) we obtain $\mathbf{u} = (\mathbf{e}_{x} + \gamma \sin{\chi} \, \mathbf{e}_{y}) \sin{(2\pi \tau)}$. It is then convenient to introduce the following signature of the drift motion: $\mathbf{s} = \lim_{n \to \infty} \int_{n}^{n+1} \mathbf{u}(\tau) d\tau$ ($n$ is a whole number). Because the particle displacement during the time period $2\pi/\Omega$ is given by $(2\pi v_{0} /\Omega) \mathbf{s}$, this quantity can be interpreted as the dimensionless drift velocity of particles in the steady state. By representing the dimensionless time in the form $\tau = n + \xi$ ($\xi \in [0,1]$), one gets $\mathbf{s} = \int_{0}^{1} \mathbf{u}_{ \mathrm{st}}(\xi) d\xi$, where $\mathbf{ u}_{ \mathrm{st}}(\xi) = [\mathbf{e}_{x} + \gamma \sin{ [\chi_{\mathrm{st}}(\xi + \phi/2\pi)]} \, \mathbf{e}_{y}] \sin{(2\pi \xi)}$. The steady-state solution of Eq.~(\ref{eq_chi}), $\chi_{\mathrm{st}} = \chi_{\mathrm{ st}} (\xi + \phi/2\pi)$, is defined as $\chi_{\mathrm{st}} = \lim_{n \to \infty} \chi(n + \xi)$ and satisfies the condition $\chi_{ \mathrm{st}} |_{1/2 +\xi} =-\chi_{\mathrm{st}} |_{\xi}$. From this it follows that while the $x$ component of the drift velocity equals zero, $s_{x} = 0$, the drift velocity along (against) the $y$ axis, which is caused by the Magnus force, is in general nonzero,
\begin{equation}
    s_{y} = 2\gamma \int_{0}^{1/2} \sin{[\chi_{
    \mathrm{st}}}(\xi)] \sin{(2\pi \xi - \phi)}
    d\xi.
    \label{s_y}
\end{equation}
It should be emphasized that $\chi_{\mathrm{st}} (\xi)$ in (\ref{s_y}) is the steady-state solution of Eq.~(\ref{eq_chi}) at $\phi =0$.

Formula (\ref{s_y}) can be rewritten as $s_{y}\! =\! -S \sin{(\phi + \delta)}$, where $S =\! 2\gamma \sqrt{I_{s}^{2} + I_{c}^{2}}$, $I_{s} = \int_{0}^{1/2} \sin{[\chi_{\mathrm{st}}}(\xi)] \sin{(2\pi\xi)} d\xi$, $I_{c} = \int_{0}^{1/2} \sin{[\chi_{\mathrm{st}}}(\xi)] \cos{(2\pi\xi)} d\xi$, and the angle $\delta$, which characterizes the rotational relaxation of particles, satisfies the condition $\tan{\delta} = \kappa$ $(\kappa = -I_{s}/I_{c})$. As seen, the drift velocity as a function of the initial phase $\phi$ lies in the interval $[-S,S]$. If the integrals $I_{s}$ and $I_{c}$ have opposite signs, then $s_{y} = \mathrm{sgn} (I_{s})S$ ($\mathrm{sgn}$ is the sign function) at $\phi = \pi/2- \arctan{|\kappa|}$ and $s_{y} = -\mathrm{sgn} (I_{s})S$ at $\phi = 3\pi/2 - \arctan{|\kappa|}$. In contrast, if $I_{s}$ and $I_{c}$ have the same sign, then $s_{y} = \mathrm{sgn} (I_{s})S$ at $\phi = 3\pi/2 + \arctan{ |\kappa|}$ and $s_{y} = -\mathrm{sgn} (I_{s})S$ at $\phi = \pi/2 + \arctan{| \kappa|}$. Thus, the initial phase $\phi$ controls both the magnitude and direction of the drift velocity.

It should be stressed that, since thermal fluctuations are ignored, the above results are valid for relatively large particles. More precisely, the core radius $a$ must exceed the superparamagnetic critical radius $a_{1}$, which ranges from a few to tens of nanometers \cite{Guim}. At the same time, because the core state is assumed to be single-domain, the core radius should be less than the critical one $a_{2}$ (in the opposite case, when $a > a_{2}$, the multi-domain state is energetically favored) \cite{Skom}. Therefore, if the condition $a \in (a_{1}, a_{2})$ holds, then the ferromagnetic core is single-domain and the magnetization dynamics is deterministic (for more details see Refs.~\cite{LDPB, DLPH}). As to the translational and rotational Brownian motions, they can be neglected if the shell radius $b$ exceeds the critical one $b_{1}$ ($b_{1}$ is usually of the order of a few hundreds of nanometers). Summarizing, we conclude that in our model the conditions $a \in (a_{1}, a_{2})$ and $b> b_{1}$ must also be satisfied. We remark in this connection that, since $a_{2}$ can be of the order of hundreds nanometers \cite{Guim}, the model parameter $\nu$ can take any value in the interval $(0,1)$.

\textit{Linear approximation.}---We start our analysis of the dependence of the drift velocity on the steady-state solution of Eq.~(\ref{eq_chi}) with the linear approximation. In this case, Eq.~(\ref{eq_chi}) reduces to the linear one $\dot{ \chi} + \alpha \chi =\dot{\psi}$, whose general solution on the interval $[0,\tau]$ is given by
\begin{equation}
    \chi(\tau) = e^{-\alpha \tau} \chi(0) +
    e^{-\alpha \tau} \int_{0}^{\tau} \dot{\psi}
    (\tau')\, e^{\alpha \tau'} d\tau'.
    \label{chi1}
\end{equation}
In the steady state, the first term on the right of (\ref{chi1}) vanishes and the second one can be transformed to yield
\begin{equation}
    \chi_{\mathrm{st}}(\xi) = \left( \Psi(\xi) -
    \frac{\Psi(1/2)}{1 + e^{\alpha/2}} \right)
    e^{-\alpha \xi}
    \label{chi_st1}
\end{equation}
with $\Psi(\xi) = \int_{0}^{\xi} \dot{\psi}(\xi')\, e^{\alpha \xi'} d\xi'$.  Therefore, replacing $\sin{[\chi_{ \mathrm{st}}} (\xi)]$ in (\ref{s_y}) by $\chi_{\mathrm{st}} (\xi)$ from (\ref{chi_st1}) and taking into account that $\int e^{-\alpha \xi} \sin{(2\pi \xi - \phi)} d\xi = - (1/2\pi) e^{-\alpha \xi}\times \cos{\beta} \cos{(2\pi \xi - \phi - \beta)}$, where $\beta = \arctan{(\alpha/ 2\pi)}$, we get
\begin{equation}
    s_{y} = \frac{\gamma}{\pi} \cos{\beta}\int_{0}^{1/2}
    \dot{\psi}(\xi) \cos{(2\pi \xi - \phi - \beta)} d\xi.
    \label{s_y1}
\end{equation}

In particular, if $\psi (\xi)= \psi_{m} \cos{ (2\pi \xi)}$ [$\psi_{m}$ is the amplitude of the azimuthal angle $\psi (\xi)$], then (\ref{chi_st1}) and (\ref{s_y1}) give $\chi_{ \mathrm{st}} (\xi) = \psi_{m} \cos{\beta} \cos{(2\pi \xi + \beta)}$ and $s_{y} = -(\gamma \psi_{m}/2) \cos{\beta} \sin{(\phi + \beta)}$, respectively. From this one finds $S = (\gamma \psi_{m}/2) \cos{\beta}$, $I_{s} = -(\psi_{m}/ 4)\sin{\beta} \cos{\beta}$, $I_{c} = (\psi_{m}/ 4) \cos^{2}{\beta}$, and $\delta = \beta$. In addition, since $I_{s} <0$ and $I_{c} >0$, the maximum drift velocity along the $y$ axis (when $s_{y} = S$) occurs at $\phi = 3\pi/2 - \beta$ and the maximum drift velocity against the $y$ axis (when $s_{y} = -S$) occurs at $\phi = \pi/2 - \beta$. An important feature of the drift velocity, which follows directly from the above expression for $s_{y}$, is that not only the magnitude of velocity, but also its direction, i.e., the sign of $\sin{(\phi + \beta)}$, depends on the parameter $\alpha$. (This property of $s_{y}$ holds also beyond the linear approximation and may be used to separate particles, see below.) Note also that the condition of applicability of the linear approximation, $|\chi_{ \mathrm{st}} (\xi)| \ll 1$, in this case reduces to $\psi_{m}/ \sqrt{1 + (\alpha/ 2\pi)^{2}} \ll 1$.

\begin{figure}
    \centering
    \includegraphics[totalheight=4cm]{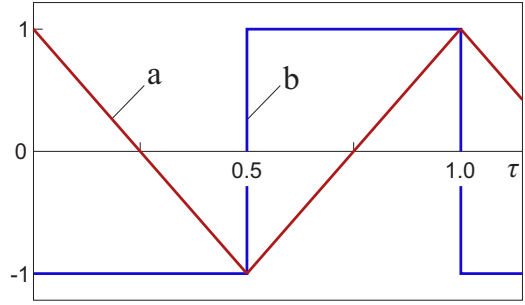}
    \caption{\label{fig1} (Color online) Plots of
    the functions $\psi(\tau) /\psi_{m}$ (a) and
    $\dot{\psi}(\tau) /4\psi_{m}$ (b) for the
    azimuthal angle (\ref{def_psi}).}
\end{figure}

\textit{Exactly solvable case.}---Next we consider the drift motion of particles occurring in the magnetic field characterized by the azimuthal angle (see Fig.~\ref{fig1})
\begin{equation}
    \psi(\tau) = \tfrac{2}{\pi} \psi_{m} \arcsin\!
    {\big[\! \cos {(2\pi \tau)}\big]}.
    \label{def_psi}
\end{equation}
The advantage of this choice of $\psi(\tau)$ is that the steady-state solution of Eq.~(\ref{eq_chi}) can be determined analytically. Indeed, since $\dot{\psi}(\tau) = \mp 4\psi_{m}$ [the signs `$-$' and `$+$' correspond to the intervals $(n, n+1/2)$ and $(n+1/2, n+1)$, respectively], Eq.~(\ref{eq_chi}) in each of these intervals can be solved by using the variable separation method. The solution depends on whether the parameter $4\psi_{m}$ is greater, less or equal to $\alpha$. Here, for illustration and verification purposes, we consider only the case with $4\psi_{m} = \alpha$, when Eq.~(\ref{eq_chi}) reduces to $d\chi/ (\sin{\chi} \pm 1) = - \alpha d\tau$ (solutions in other cases will be presented elsewhere). Using the standard integral $\int (\sin{x} \pm 1)^{-1}dx = \pm \tan{(x/2 \mp \pi/4)} + c$ ($c$ is an integration constant) and the continuity conditions for $\chi(n+\xi)$, $\chi(n-0) = \chi(n+0)$ and $\chi(n+1/2-0) = \chi(n+1/2+0)$, we find
\begin{equation}
    \tan\!{\left[ \tfrac{1}{2}\chi(n+\xi) -
    \tfrac{\pi}{4} \right]} = - \alpha \xi +
    \tan\!{\left[\tfrac {1}{2}\chi(n) -
    \tfrac{\pi}{4}\right]}
    \label{chi2}
\end{equation}
for $\xi \in [0, 1/2]$ and
\begin{equation}
    \tan\!{\left[ \tfrac{1}{2}\chi(n+\xi) +
    \tfrac{\pi}{4} \right]} = \alpha\!\left(\xi -
    \tfrac{1}{2}\right) + \tan\!{\left[\tfrac{1}{2} \chi\!\left(n+\tfrac{1}{2}\right) + \tfrac
    {\pi}{4} \right]}
    \label{chi3}
\end{equation}
for $\xi \in [1/2, 1]$. Then, since $\lim_{n \to \infty} \chi(n + \xi) = \chi_{\mathrm{st}}(\xi)$, from Eqs.~(\ref{chi2}) and (\ref{chi3}) one obtains $\chi_{\mathrm{st}}(0) = \arctan{ (\alpha/4)}$, $\chi_{\mathrm{st}}(1/2 + \xi) = - \chi_{\mathrm {st}}(\xi)$, and
\begin{equation}
    \tan\!{\left[ \tfrac{1}{2}\chi_{\mathrm{st}}
    (\xi) - \tfrac{\pi}{4} \right]} = - \alpha
    \xi + \tan\!{\left[ \tfrac{1}{2}\chi_{\mathrm{st}}(0) -
    \tfrac{\pi}{4} \right]}
    \label{eq_chi_st}
\end{equation}
for $\xi \in [0, 1/2]$. Finally, introducing the designation $q(\xi) = \alpha\xi - \alpha/4 + \sqrt{1 + (\alpha/4)^{2}}$ and using standard trigono\-metric identities, Eq.~(\ref{eq_chi_st}) can be represented as $\sin{[\chi_{ \mathrm{st}}(\xi)]} = [1 - q^{2}(\xi)]/[1 + q^{2}(\xi)]$, which simplifies the drift velocity (\ref{s_y}) at $4\psi_{m} = \alpha$ to
\begin{equation}
    s_{y} = -\frac{2\gamma}{\pi}\cos{\phi} +
    4\gamma \int_{0}^{1/2} \frac{\sin{(2\pi\xi -
    \phi)}}{1 + q^{2}(\xi)} d\xi.
    \label{s_y2}
\end{equation}

Using the last result, one straightforwardly obtains $s_{y} \approx - (\alpha \gamma/\pi^{2}) \sin{\phi}$ for $\alpha/4 \ll 1$ and $s_{y} \approx - (2\gamma/ \pi) \cos{\phi}$ for $\alpha/4 \gg 1$. From this it follows that the dependence of the drift velocity on the parameter $\alpha$ is nonmonotonic. Moreover, if $\phi \in (\pi/2, \pi)$ or $\phi \in (3\pi/2, 2\pi)$, then the sign of $s_{y}$, i.e., the drift direction, also depends on $\alpha$.

\begin{figure}
    \centering
    \includegraphics[totalheight=4cm]{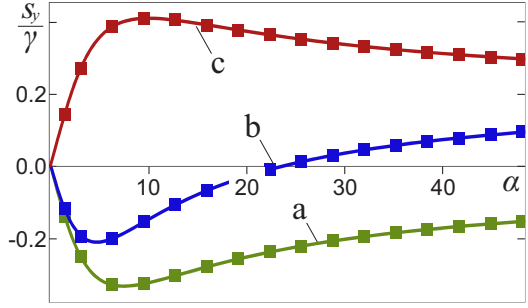}
    \caption{\label{fig2} (Color online) Theoretical
    (solid lines) and numerical (square symbols)
    dependencies of the reduced drift velocity
    $s_{y}/ \gamma$ on the parameter $\alpha$ at
    $4\psi_{m} =\alpha$. The initial phase is chosen
    to be $\phi = 1.4$ (a), $\phi = 1.9$ (b), and
    $\phi = 4.4$ (c).}
\end{figure}

\textit{Numerical results.}---In order to verify the theoretical predictions, we first calculated the reduced drift velocity $s_{y} /\gamma$, which is defined by (\ref{s_y2}), as a function of the parameter $\alpha$ for different values of the initial phase $\phi$ (see Fig.~\ref{fig2}, solid lines; $\phi$ is measured in radians). Then, using Eq.~(\ref{eq_chi}) whose right-hand side is determined by the azimuthal angle (\ref{def_psi}) with $4\psi_{m} =\alpha$, we found numerically the steady-state solution of this equation, $\chi_{\mathrm{ st}}(\xi)$ ($0 \leq \xi \leq 1/2$), for a set of the parameter $\alpha$. Finally, calculating the integral in (\ref{s_y}) for the same values of $\phi$ as in (\ref{s_y2}), we determined $s_{y} /\gamma$ for the chosen values of the parameters $\alpha$ and $\phi$ (see Fig.~\ref{fig2}, square symbols). As seen, our analytical results are in complete agreement with the numerical ones derived from the basic equations.

\begin{figure}
    \centering
    \includegraphics[totalheight=4cm]{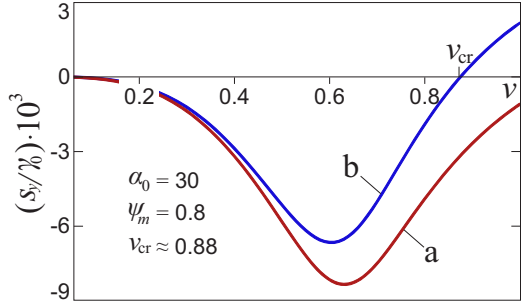}
    \caption{\label{fig3} (Color online) Reduced
    drift velocity $s_{y}/ \gamma_{0}$ as a function
    of the core-shell ratio $\nu$ for $\phi= 0.55$ (a)
    and $\phi = 0.6$ (b). The other parameters are
    shown in the figure.}
\end{figure}

The numerical analysis, which includes the case with $4\psi_{m} \neq \alpha$, shows that the most striking feature of the drift velocity $s_{y}$ is that the change of the parameter $\alpha$ can lead (if the initial phase $\phi$ and azimuthal angle amplitude $\psi_{m}$ are chosen appropriately) to the change of the sign of $s_{y}$. Since $\alpha$ depends on the core-shell ratio $\nu$, one can therefore distinguish two different behaviors of $s_{y}$ on $\nu$. The first is realized if the equation $s_{y}=0$ with respect to $\nu$ has no solution. In this case, the sign of $s_{y}$ is the same for all particles, and so their unidirectional drift is only possible (see curve a in Fig.~\ref{fig3}). The second occurs if the equation $s_{y}=0$ has a solution $\nu =\nu_{ \mathrm{cr}}$. In contrast to the previous case, the particles with $\nu <\nu_{ \mathrm{cr}}$ and $\nu >\nu_{ \mathrm{cr}}$ are now expected to drift in opposite directions (see curve b in Fig.~\ref{fig3}). This bidirectional drift may exist in suspensions of particles with different values of $\nu$; if the suspended particles are identical, then only unidirectional drift can be induced. From a physical point of view, the drift is caused by coherent translational and rotational oscillations of particles. Since, according to (\ref{s_y}), $s_{y}|_{\phi\pm \pi} = -s_{y}|_{ \phi}$, for each parameter $\alpha = \alpha'$ there is an initial phase (coherence parameter) $\phi= \phi'$ such that $s_{y} |_{\phi'} =0$. Therefore, if for suspended core-shell particles the condition $\alpha_{0} < \alpha'$ holds, then at $\phi= \phi'$ coherent oscillations of all particles occur in such a way that only unidirectional drift is realized. Otherwise, if $\alpha_{0} > \alpha'$, the bidirectional drift is expected.

The bidirectional drift of core-shell particles might turn out to be useful for the development of a novel technique for their separation in suspensions. To illustrate this, let us consider a dilute suspension of particles whose core-shell ratio $\nu$ is randomly distributed. In equilibrium, the concentration of particles with different values of $\nu$ is spatially homogeneous. In contrast, the concentration of particles subjected to the periodic force and magnetic field is expected to be inhomogeneous due to their different drift velocity. If the conditions for bidirectional drift hold, then the separation of particles with respect to their core-shell ratio should occur in opposite directions along the preferred axis (axis $y$). Specifically, the particles with $\nu <\nu_{ \mathrm{cr}}$ are concentrated mainly near one side of the suspension sample, and the particles with $\nu >\nu_{ \mathrm{cr}}$ near the opposite side. It is worthwhile to emphasize that the magnitude of $\nu_{ \mathrm{cr}}$ can be changed by changing the initial phase $\phi$ of the magnetic field.

The separation time, i.e., the average time of the separation process, can be estimated as $t_{\mathrm{sep}} = l /\max{v_{ \mathrm{dr}}}$, where $l$ is the sample size in the $y$ direction and $\max{v_{ \mathrm{dr}}} = v_{0} \max{|s_{y}|} \sim v_{0} \gamma_{0}$ is the maximum drift velocity of particles. In particular, for $\mathrm{SmCo_{5}}$ particles suspended in water at room temperature and subjected to the oscillating magnetic field one gets $\alpha_{0} \approx 8.2 \cdot 10^{3}$ and $\gamma_{0} \approx 0.22$ as $M= 1.4\cdot 10^{3}\; \mathrm{emu /cm^{3}}$, $\rho = 1\; \mathrm{g/cm^{3}}$, $\eta =8.9\cdot 10^{-3}\; \mathrm{P}$, $H_{m} = 5\cdot 10^{2}\; \mathrm{Oe}$, $\Omega =10^{4}\; \mathrm{ rad/s}$, and $a=b = 3\cdot 10^{-5}\; \mathrm{cm}$. Assuming also that $l = 1\; \mathrm{cm}$ and $v_{0} = 10^{-2}\; \mathrm{cm/s}$ (for comparison, if particles move under the influence of the magnetic field with a gradient of $10^{3}\; \mathrm{Oe/cm}$, then $f_{0} \approx 1.6 \cdot 10^{-7}\; \mathrm{g \cdot cm/s^{2}}$ and so $v_{0} \approx 3.2 \cdot 10^{-2}\; \mathrm{cm/s}$), we obtain $t_{\mathrm{ sep}} \sim 7.6\; \mathrm{min}$. Since the $\mathrm{SmCo_{5}}$  particles of a given size are single domain \cite{GGI} and the Reynolds numbers are small enough, $\mathrm{Re}_{t} \approx 3.4 \cdot 10^{-5}$ and $\mathrm{Re}_{r} \approx 1.0 \cdot 10^{-3}$ (we take $v_{m} = v_{0}$ and $\omega_{m} = \Omega$), these theoretical estimates seem to be quite reliable.

\textit{Conclusions.}---We have shown analytically and numerically that the joint action of a time-periodic force and a nonuniformly rotating magnetic field on ferromagnetic particles suspended in a viscous fluid can induce their drift in a preferred  direction caused by the Magnus force. We have found that in suspensions, which are composed of particles with single-domain cores, nonmagnetic shells and different core-shell ratios, two types of drift, unidirectional and bidirectional, can be realized by an appropriate choice of the magnetic field initial phase. It is also argued that the bidirectional drift can be used to separate suspended particles by core-shell ratio.

This work was supported by the Ministry of Education and Science of Ukraine under Grant No.\ 0116U002622.

\end{document}